\documentclass[12pt, preprint]{aastex}
\usepackage{natbib,graphics}
\shorttitle{{\em Fermi} Constraints on Decaying Dark Matter}

\shortauthors{Dugger, Jeltema and Profumo}

\begin{document}

\title{Constraints on Decaying Dark Matter from  \\ {\em Fermi} Observations of Nearby  Galaxies and Clusters}

\author{Leanna Dugger\altaffilmark{1,2} Tesla E. Jeltema\altaffilmark{3} and Stefano Profumo\altaffilmark{1}}

\altaffiltext{3}{UCO/Lick Observatories, 1156 High St., Santa Cruz, CA 95064}
\altaffiltext{2}{Department of Physics and Santa Cruz Institute for Particle Physics,  University of California, 1156 High St., Santa Cruz, CA 95064, USA}
\altaffiltext{1}{University of California, Berkeley}

\begin{abstract}
\noindent We analyze the impact of {\em Fermi} gamma-ray observations (primarily non-detections) of selected nearby galaxies, including dwarf spheroidals, and of clusters of galaxies on decaying dark matter models. We show that the fact that galaxy clusters do not shine in gamma rays puts the most stringent limits available to-date on the lifetime of dark matter particles for a wide range of particle masses and decay final states. In particular, our results put strong constraints on the possibility of ascribing to decaying dark matter both the increasing positron fraction reported by PAMELA  and the high-energy feature in the electron-positron spectrum measured by {\em Fermi}. Observations of nearby dwarf galaxies and of the Andromeda Galaxy (M31) do not provide as strong limits as those from galaxy clusters, while still improving on previous constraints in some cases.
\end{abstract}


\section{INTRODUCTION}

A large variety of independent astronomical observations have revealed that most of the mass in the universe is some form of non-baryonic, non-luminous, cold dark matter of as yet unknown composition.  Understanding the fundamental nature of dark matter is one of the biggest outstanding problems in cosmology and particle physics \citep[for a comprehensive review, see e.g.][]{bertonebook}.  Compelling particle candidates for the dark matter, collectively termed weakly interacting massive particles (WIMPs), exist in several well-motivated theoretical extensions of the Standard Model of particle physics like supersymmetry \citep[lightest supersymmetric particle, for a review see e.g.][]{1996PhR...267..195J} and Universal Extra Dimensions \citep[lightest Kaluza-Klein particle, for a review see e.g.][]{2007PhR...453...29H}.  A signal from dark matter could be detected in astronomical observations from WIMP pair annihilation or decay.  For example, WIMP annihilation or decay generically leads to the production of gamma-rays (as well as of other high energy particles), potentially at a detectable level with current telescopes \citep{2005PhR...405..279B}.  The unparalleled sensitivity of the {\em Fermi} Large Area Telescope (LAT) at GeV energies makes it an excellent instrument to look for this gamma-ray signal \citep[e.g.][]{2008JCAP...07..013B}.  In addition, the recent detection by PAMELA of a positron excess relative to cosmic-ray secondary production in the Galaxy \citep{2009Natur.458..607A} and a gentle but clear feature in the {\em Fermi}-LAT positron-electron ($e^+-e^-$) spectrum \citep{2009PhRvL.102r1101A} has led to intense discussion of the possibility that one or both of these signals are due to dark matter.  If true, we expect a corresponding signal in gamma-rays stemming from both the annihilation or decay event itself, as well as from the secondary inverse Compton (IC) up-scattering of background radiation by the high energy $e^+$ and $e^-$ produced. 

As far as dark matter annihilation is concerned, first year {\em Fermi} observations of dwarf spheroidal galaxies, clusters of galaxies, the extragalactic gamma-ray background, and searches for gamma-ray lines have placed some of the strongest constraints yet on the possible WIMP pair annihilation cross-section and have already excluded some annihilation models that might explain the PAMELA positron excess \citep{dwarfdm, clusdm, 2010JCAP...04..014A, 2010PhRvL.104i1302A}.  In addition, several authors have recently pointed out that gamma-ray data (including early {\em Fermi} results) also put constraints on the dark matter decay explanation of the PAMELA data, excluding some, but not all, of the relevant parameter space \citep[e.g.][]{2010arXiv1008.3636I, 2010JCAP...03..014P, 2010NuPhB.840..284C, 2010JCAP...01..023C, 2010NuPhB.831..178M, 2010JCAP...06..027Z, 2010ApJ...712L..53P}.  These studies typically utilize {\em Fermi}-LAT measurements of the diffuse Galactic and isotropic gamma-ray emission or {\em Fermi} sky maps and require that the expected decay signal from the Milky Way halo or the summed contribution from extragalactic dark matter not overproduce the observed gamma-ray fluxes.   Limits on the dark matter lifetime from the observation of dwarf galaxies with atmospheric Cherenkov telescopes have also been derived in \cite{2009PhRvD..80b3506E}.  Interestingly, these analyses indicate that current gamma-ray constraints on the decaying dark matter scenario are generically weaker than those on annihilating dark matter, with respect to the regions of parameter space providing an explanation to the cosmic-ray electron-positron anomalies. Moreover, as we review in the next section, several theoretical arguments imply that dark matter lifetimes in the range needed to explain the cosmic-ray anomalies are at least plausible, specifically in the context of supersymmetric grand unification \citep{2009PhRvD..79j5022A}. 

Beyond the interest in scenarios where decaying dark matter might be related to the cosmic-ray anomalies, several particle physics frameworks feature dark matter candidates with a long, but finite, lifetime. We provide an overview of some of these models in the next section. The present study targets this general class of theories as well, and we consider here decay final states that can be regarded as representative for a wide range of scenarios.
Searches for dark matter decay with gamma ray observations are therefore not only phenomenologically motivated, as they can test the dark matter interpretation of the observed cosmic-ray anomalies, but they also probe theoretically compelling extensions to the Standard Model of particle physics.

While already significant, much can be done to improve the current limits on decaying dark matter models.  In this work, we show that novel and significant constraints can be placed on the dark matter decay lifetime for both quark-antiquark decay final states as well as for leptonic final states giving a good fit the cosmic-ray anomalies using {\em Fermi}-LAT observations of isolated dark matter dominated objects like nearby clusters of galaxies, dwarf spheroidal galaxies, and M31.  In fact, the simulations of \cite{2010arXiv1007.3469C} which compared the gamma-ray flux from dark matter annihilation and decay in nearby large-scale structures as reconstructed from constrained cosmological simulations predict that nearby clusters and filaments are good targets for searches for and constraints on a signal from dark matter decay, and here we derive such constraints with recent \textit{Fermi} observations.  In \S2, we discuss and introduce selected dark matter decay models, in \S3 we discuss the selection of targets and the modeling of their dark matter content, and in \S4 we present our results on the constraints on the decay lifetime.

\section{MODELS OF DECAYING DARK MATTER}

While stable on the scale of cosmological structure formation, the particle making up the universal dark matter might well have a finite lifetime, $\tau$. If this is the case, and if the dark matter particle decays into (visible) Standard Models particles, the decay process can be {\em indirectly} detected. Depending on the particle lifetime, non-thermal secondary particles resulting from dark matter decay can affect key ``episodes'' in the early universe such as the synthesis of light elements or the decoupling of the cosmic microwave background (CMB). In turn, light elemental abundances and the CMB spectrum can be used to set constraints on particle models where the dark matter is unstable \citep[see e.g.][]{marcspaper}. 

There is no shortage of particle models where the universal dark matter is unstable, including unstable supersymmetric particles, moduli as dark matter particles, super-heavy dark matter candidates, axinos and sterile neutrinos \citep{marcspaper}. Interestingly, particle candidates that would otherwise be completely un-detectable, such as gravitinos in supersymmetric extensions of the Standard Model, can in principle provide indirect clues on their nature if they decay \citep{1991PhLB..266..382B}. Among the decay particle debris, gamma rays and antimatter (such as energetic cosmic-ray positrons, anti-protons and anti-deuterons) are ideal signals for indirect dark matter detection. As such, gamma-ray and cosmic-ray antimatter data have been extensively used to set constraints on the lifetime of meta-stable particle dark matter candidates \citep{bertonebook}.

The recent detection of an exceptionally large flux of high-energy cosmic-ray positrons ($E_{e^+}\gtrsim 10$ GeV), compared to conventional models of Galactic secondary positron production, reported by the PAMELA space-borne experiment \citep{2009Natur.458..607A} prompted great interest in possible non-standard sources. Another recently launched satellite experiment, {\em Fermi}-LAT, also showed the existence of a gentle but clear feature in the spectrum of cosmic-ray electrons-plus-positrons at energies of several hundreds of GeV, especially when supplemented with even higher energy data from the H.E.S.S. atmospheric Cherenkov Telescope \cite{hessepem}. While possibly pointing to nearby astrophysical sources such as mature (characteristic age $\gtrsim10^5$ yr) pulsars \citep[see e.g.][]{psr1,psr2,psr3,psr4}, the feature detected by {\em Fermi}-LAT and the rise in the PAMELA positron fraction might share a common, more exotic origin: Galactic dark matter \citep{2009Natur.458..607A}. A large number of studies have, in fact, demonstrated that particle models can be constructed to accommodate both the positron rise and the {\em Fermi} ``feature'' (for short, we will hereafter dub these two as the ``cosmic-ray anomalies''), invoking either the pair-annihilation or the decay of dark matter particles \citep[see e.g.][]{2010JCAP...03..014P}. 

A dark matter annihilation interpretation of the cosmic-ray anomalies requires a very large pair-annihilation cross section, one or more orders of magnitude larger (depending on the dark matter particle mass) than the one that would explain via thermal decoupling the observed universal matter density \citep{bergstrom}. The attractiveness of the thermal relic dark matter scenario must therefore be traded-off, within this scenario, for large annihilation rates in today's cold universe, unless a very strong velocity dependence in the annihilation cross-section is invoked, \citep[see e.g.][]{2009PhRvD..79a5014A, 2009PhRvD..79i5009I, 2009PhRvD..80h1302C}. Dark matter decay as the source of the cosmic-ray lepton anomalies, on the other hand, requires the fine-tuning of the decay time-scale $\tau$, but is otherwise a phenomenologically perfectly viable scenario. The required lifetimes are on the order of a few $10^{26}$ s, for particle masses in the TeV range.

In the broader picture of particle model-building, several scenarios predict that the symmetry that preserves the stability of the dark matter particle on scales much larger than the age of the universe is violated by sufficiently suppressed higher-dimension operators (i.e. operators with mass-dimension greater than 4) that mediate the subsequent, slow particle decay. Similarly to the generic prediction of proton decay, in the context of supersymmetric grand unification, operators with mass dimension 6 are expected to make supersymmetric dark matter unstable, with a natural time-scale
\begin{equation}\label{eq:gut}
\tau\sim8\pi M_{\rm GUT}^4/m_{\rm DM}^5,
\end{equation}
where $M_{\rm GUT}$ is the energy scale of grand unification interactions, as recently pointed out by \cite{2009JCAP...01..043N} and, subsequently, by \cite{2009PhRvD..79j5022A}. Interestingly this yields the prediction $$\tau\sim10^{27}{\rm sec}\left(\frac{\rm TeV}{m_{\rm DM}}\right)^5\left(\frac{M_{\rm GUT}}{2\times 10^{16}\ {\rm GeV}}\right)^4$$ which is precisely in the ballpark needed to explain the cosmic-ray anomalies. Notice that GUT-suppressed dimension-5 operators would produce lifetimes on the order of a second \citep{2009PhRvD..79j5022A}, and must thus be suppressed to avoid dark matter decaying long before structure formation.

In addition to grand unification, another compelling reason to expect dark matter to decay with a lifetime on the order of $10^{26}-10^{27}$ s is provided by models, such as those elucidated in the analysis of \cite{2009JCAP...02..021I}, where the lightest supersymmetric particle is a gravitino and $R$-parity is (very weakly) broken \cite[see also][]{2007JCAP...11..003B, 2008PhRvL.100f1301I}. In this class of models, not only is the supersymmetric gravitino problem naturally solved, but the correct baryon asymmetry can naturally be generated by $R$-parity violating interactions via leptogenesis. Another scenario that might in principle combine the origin of both the baryonic matter and the dark matter via leptogenesis, and provide a mechanism to explain neutrino masses and mixing is that of sterile neutrinos \citep[for a review see e.g. ][]{2009ARNPS..59..191B}. Recent analyses also pointed to hidden-sector gauge bosons and gauginos as natural particle frameworks where TeV-scale particle dark matter decays with lifetimes on the order of $10^{26}$ s. A scenario featuring leptonic decay final states and a sufficiently long-lived dark matter as a consequence of a global symmetry broken by instanton operators originating from a non-Abelian gauge group in the dark sector was also discussed in \cite{2010PhRvD..82e5028C}.

If the decay of particle dark matter is to explain the cosmic-ray lepton anomalies, the required dominant decay final state must be close to one (or more) muon pairs, as shown e.g. in \cite{2010JCAP...03..014P}. Nevertheless, among many of the scenarios mentioned above, the decay products will likely follow hadronization chains that are typically well reproduced in their major features by decay final states such as a quark-antiquark pair (even though for instance in the scenario of \cite{2009JCAP...02..021I} the final state might be that resulting from $W$ or $Z$ decay, the hadronic decay modes would still produce a gamma-ray spectrum qualitatively similar to that of a quark-antiquark pair; see also \cite{2010arXiv1009.0224C} for a discussion of weak corrections to annihilation or decay final states). For this reason, in the present study we will focus on two specific decay final states: $\mu^+\mu^-$ as representative of those scenarios that offer a particle dark matter decay interpretation to the cosmic-ray anomalies; and $b\bar b$ for more conventional decaying dark matter scenarios.

For these two decaying dark matter final states, we will consider in detail the produced gamma-ray spectrum as a function of the dark matter particle mass, and model this spectrum with the DMFIT package \citep{2008JCAP...11..003J} as implemented in the {\em Fermi}-LAT data analysis pipeline\footnote{http://fermi.gsfc.nasa.gov/ssc/}. The consideration of different decay final states will be coupled to the choice of the best targets, as one particular target does not necessarily give the strongest constraints for all dark matter models. An important effect, as shown in \cite{dracoullio}, \cite{profumojeltemadsph}, and \cite{2010ApJ...712..147A}, is the diffusion of the high-energy electrons and positrons produced by dark matter decay. For sufficiently massive dark matter candidates, in fact, these particles copiously up-scatter low-energy background photons (inverse Compton -- IC -- emission) into gamma rays detectable by the LAT \citep{2010JCAP...06..027Z,2010arXiv1008.1801Z}. The IC photon flux is however strongly dependent upon the electron diffusion and leakage out of the observed target, an effect that impacts both the spectrum and the predicted flux-intensity. We discuss this and other target-dependent considerations in the next section. 

\section{TARGETS FOR DARK MATTER DECAY SEARCHES}

A number of astrophysical targets are potentially good sources to search for a signal from dark matter annihilation or decay, including the Galactic center, the Milky Way halo, nearby galaxies and dwarf spheroidal galaxies, nearby clusters of galaxies, and the summed contribution from unresolved extragalactic dark matter \citep[for a review in the context of the LAT performance, see e.g.][]{2008JCAP...07..013B}.  Underpinning which is the ``best'' target depends, however, on the type of signal (annihilation or decay), on our understanding of the dark matter density in different objects, on the gamma-ray backgrounds, and on the dominant gamma-ray emission mechanism for a given dark matter particle model (for instance prompt gamma-ray production versus secondary IC emission).  Extragalactic objects like clusters of galaxies and Local Group dwarf galaxies are particularly attractive targets for dark matter searches as they are highly dark matter dominated, isolated objects, and many of them lie in regions of low diffuse gamma-ray background, being located at high Galactic latitudes.  In addition, with the exception of M31,  the objects we discuss do not themselves host detectable gamma-ray emission, making them clean targets of study to constrain emission from dark matter.

Previous studies looking at constraints on dark matter models using {\em Fermi} observations of dwarfs and clusters have focused on dark matter annihilation \citep{clusdm,dwarfdm}, while constraints on dark matter decay with \textit{Fermi} have primarily compared the observed diffuse gamma-ray backgrounds to the total summed emission expected from dark matter decay in the Milky Way halo or from unresolved extragalactic dark matter \citep{2010arXiv1008.3636I, 2010JCAP...03..014P, 2010NuPhB.840..284C, 2010JCAP...01..023C, 2010NuPhB.831..178M, 2010JCAP...06..027Z, 2010ApJ...712L..53P}. In this paper, we constrain dark matter decay models using, instead, isolated extra-galactic dark matter sources, specifically nearby clusters, dwarf spheroidal satellites of the Milky Way, and the nearby galaxy M31 (the Andromeda galaxy).  
In some cases, these targets provide significantly stronger constraints on the dark matter decay lifetime than have been previously published, particularly for the case of nearby clusters \citep[see also the simulations of][]{2010arXiv1007.3469C}, which we discuss in the next section.

\subsection{CLUSTERS OF GALAXIES}

Clusters of galaxies are the most massive collapsed objects in the present-day Universe and are highly dark matter dominated.  Apart from the central bright active galactic nuclei in a couple of clusters, clusters have thus far not been detected in gamma-rays, a fact that allows one to place limits on their possible dark matter related emission \citep{clusdm,cluscr,hesscoma,magicperseus}. We consider here the same sample of six, nearby clusters of galaxies as we studied in \cite{clusdm}.  These clusters are among the best candidates for dark matter searches and have published upper limits on their potential gamma-ray emission from dark matter (annihilation) based on the first 11 months of {\em Fermi}-LAT observations for two illustrative (annihilation) final states, $b\bar b$ and $\mu^+\mu^-$, and for a grid of dark matter particle masses.  
Remarkably, the expected gamma-ray spectral shape from dark matter decay for a given particle mass is the same as the gamma-ray spectrum from annihilation for a particle of half the mass with the same assumed final state \citep[although certain final states can only be available to decay processes, such as e.g. $W$ plus charged leptons, see e.g.][]{2010JCAP...07..023P}, and we use the corresponding flux upper limits from \cite{clusdm} to take into account the spectral dependence of the broad-band {\em Fermi}-LAT upper limits.  Clusters are particularly good targets when considering decay to leptons, as is the case for models fitting the cosmic-ray anomalies, because in this case much of the gamma-ray emission stems from IC scattering of background radiation by the energetic $e^+e^-$ produced \citep{clusdm, coma}.  In clusters, $e^+e^-$ lose energy through IC much faster than they can diffuse out of the system, unlike in smaller systems like dwarf galaxies \citep{profumojeltemadsph, dracoullio, coma}.

The {\em Fermi}-LAT effective area is a strong function of energy and increases by nearly an order of magnitude from 100 MeV to 1 GeV, increasing more slowly at energies above 1 GeV\footnote{http://www-glast.slac.stanford.edu/software/IS/glast\_lat\_performance.htm}.  In addition, the LAT PSF decreases with increasing energy (effectively reducing the background for detection) and the gamma-ray backgrounds are relatively soft.  Given the expected spectral shape for the dark matter decay models we consider, these facts mean that the tightest constraints come from  the {\em Fermi} data around $\sim 1-2$ GeV \citep[for more detail see e.g.][]{clusdm, cluscr}\footnote{see also the LAT point source sensitivity for an assumed power law spectrum here: http://www-glast.slac.stanford.edu/software/IS/glast\_lat\_performance.htm}.  At these energies the {\em Fermi}-LAT Point-Spread Function (PSF) is around one degree, a radius which contains most of the virial mass of the clusters we consider.  Following \cite{clusdm}, we assume an NFW profile for the cluster dark matter density distribution with cluster masses taken from \cite{re02} and the concentration-mass relation of \cite{conc}.  The normalization of the expected gamma-ray flux from dark matter decay $J_d$ is

\begin{equation}
J_d\equiv\int_{\Delta\Omega} {\rm d}\Omega\int_{\rm l.o.s.}\rho_{\rm DM}(l){\rm d}l(\psi),
\end{equation} 
where we integrate the dark matter density $\rho_{\rm DM}$ along the line of sight (l.o.s.) in the direction $\psi$ over a solid angle $\Delta\Omega$ corresponding to a radius of one degree.  The $J_d$ values for our cluster sample are listed in Table 1 along with the masses, radii, and concentrations we employ.  The choice of density profile and integration radius only matters at the level of a factor of a few in the derived $J_d$ since, as already mentioned, the size of the {\em Fermi} PSF includes much of the cluster virial mass (see column 6 of Table 1) and since compared to annihilation, which is proportional to density squared, decay is relatively insensitive to the inner dark matter density profile.

\subsection{DWARF SPHEROIDAL GALAXIES}

Local Group dwarf spheroidal galaxies are very nearby objects which, in some cases, are even more dark matter dominated than clusters \citep{wolf}.  Similar to clusters, dwarf spheroidal galaxies have not been detected in gamma-rays \citep{dwarfdm}. \cite{dwarfdm} places limits on dark matter annihilation models based on the non-detection of eight dwarf spheroidal galaxies in the first 11 months of {\em Fermi}-LAT data; we will consider this sample of eight dwarfs in terms of limits on dark matter decay employing again the appropriate published {\em Fermi}-LAT flux limits for assumed $b\bar b$ and $\mu^+\mu^-$ final states and a grid of particle masses.

For dwarf spheroidal galaxies, the dark matter halo mass can be estimated from the line-of-sight stellar velocity dispersion.  Dwarf masses are most strongly constrained near the half light radius as the mass at the half light radius, $M_{1/2}$, is relatively insensitive to the unknown stellar velocity dispersion anisotropy \citep{wolf} \citep[see however also][]{2010arXiv1009.1813A}.  Outside of this radius the mass and dark matter density profile are less well constrained \citep[though the use of cold dark matter (CDM) priors reduces this uncertainty][]{martinez}.  The half light radii of the dwarf spheroidal galaxies we consider are smaller than the {\em Fermi}-LAT PSF at our energies of interest (see column 5 of Table 2), so we will conservatively consider $M_{1/2}$, taken from \cite{wolf}, as the lower limit on the dark matter mass probed by the {\em Fermi} observations.  We list the corresponding values of $J_d$ as well as $M_{1/2}$ and $r_{1/2}$ in Table 2.  For comparison, in Figure \ref{fig:dwarfsys} we show the constraints we would get for a range of plausible dwarf virial masses assuming the density profile given in Figure 3 of \cite{wolf} and taking the corresponding mass within one degree.

\subsection{M 31}

As the nearest large galaxy, M31 (the Andromeda galaxy) is also an excellent target for dark matter decay searches, as envisioned e.g.~for the case of sterile neutrinos in \cite{sterilenu}.  A significant detection of M31 has recently been reported in two years of {\em Fermi}-LAT data \citep{fermim31, 2010arXiv1008.2537O}.  The gamma-ray luminosity of M31 is consistent with predictions for the luminosity from cosmic ray collisions with the ISM and is likely not due to dark matter.  We will conservatively take the detected flux plus $2 \sigma$ \citep{fermim31} as the upper limit on the possible dark matter induced gamma-ray emission from M31 in order to place limits on the dark matter decay lifetime, though in all likelihood any dark matter signal from M31 is significantly fainter.  We consider both the flux resulting from the assumption that the gamma-ray emission is point-like and from the assumption of an extended source with emission following the M31 infra-red surface brightness. To find $J_d$, we employ the NFW dark matter density profile from \cite{2006MNRAS.366..996G} based on fits to surface brightness, velocity dispersion, and rotation curve data (see Table 1). We also consider the isothermal profile of \cite{2001MNRAS.323...13K}, for comparison.

\section{CONSTRAINTS ON DECAYING DARK MATTER MODELS}

In this section we present our results on the limits on the decay lifetime for dark matter particles as a function of mass from gamma-ray observations of clusters (\S\ref{sec:clusters}), of dwarf spheroidal galaxies (\S\ref{sec:dsph}) and of M31 (\S\ref{sec:m31}). As a rule of thumb, it is simple to give an estimate for the expected {\em Fermi}-LAT limits for the case of a standard decaying WIMP:
\begin{equation}
\tau\lesssim 8\times 10^{25}\ {\rm sec}\left(\frac{10^{-10}\ {\rm cm}^{-2}{\rm sec}^{-1}}{\phi_\gamma}\right)\left(\frac{J_d}{10^{18} {\rm GeV}{\rm cm}^{-2}}\right)\left(\frac{100\ {\rm GeV}}{m_{\rm DM}}\right)\left(\frac{N_\gamma}{10}\right)
\end{equation}
where $\phi_\gamma$ is the limit on the gamma-ray flux from the observed target, and $N_\gamma$ the integrated number of photons per decay within the relevant {\em Fermi}-LAT energy range (this number depends on the decay final state as well as on the decaying particle mass $m_{\rm DM}$).

\subsection{CLUSTERS OF GALAXIES}\label{sec:clusters}

Figure 1 shows the limits we derive for the possible dark matter decay lifetime assuming a $b\bar b$ final state (left panel) and $\mu^+\mu^-$ final state (right panel) derived from the 95\% confidence level upper limits on the gamma-ray flux from nearby clusters of galaxies presented in \cite{clusdm} for 11 months of {\em Fermi}-LAT observations.  For comparison, we show previous constraints on the decay lifetime derived from {\em Fermi} all-sky maps and considering the dark matter contribution from both the Milky Way halo and unresolved extragalactic dark matter from \cite{2010JCAP...06..027Z} \citep[see also][]{2010arXiv1008.3636I, 2010JCAP...03..014P, 2010NuPhB.840..284C, 2010JCAP...01..023C, 2010NuPhB.831..178M, 2010ApJ...712L..53P}.  

The shape of the constraints on the lifetime versus mass can be understood by considering the combined effect of how the gamma-ray flux depends on the decaying particle mass and of how the flux sensitivity depends on the resulting gamma-ray spectrum. Increasing the particle mass $m_{\rm DM}$, the particle number density, and hence the number of decays, decreases as $n_{\rm DM}\propto1/m_{\rm DM}$. However, the number of photons per decay {\em increases} with $m_{\rm DM}$ -- more energy density is dumped into the decay products overall as the decaying particle mass is increased.  The photons relevant to derive constraints from {\em Fermi}-LAT data, however, are restricted to a specific energy range. Particularly when secondary photons from Inverse Compton scattering are included, deriving a simple functional form for the integrated number of photons $N_{\gamma}$ within the {\em Fermi}-LAT energy range from the decay of a particle of mass $m_{\rm DM}$ versus the particle mass is non-trivial. Suffice it to say that, neglecting IC photons $N_\gamma$ grows with $m_{\rm DM}$ slower than linearly. This implies that, when using a {\em fixed} flux upper limit (as we do for the case of M31, see below) we expect the limits on the dark matter lifetime to decrease as $m_{\rm DM}^{-\alpha}$, with $0<\alpha<1$.

A more subtle effect occurs if the flux limits are calculated themselves as a function of mass, as is done in the dedicated {\em Fermi}-LAT data analyses presented in \cite{clusdm} and \cite{dwarfdm} where the simulated gamma-ray spectra from annihilation of particles into given final states were employed. The flux upper limits {\em improve} (i.e. decrease) with the particle mass, since most of the photons are produced at increasingly large energies where the instrumental point-spread function and effective area improve \citep[see e.g. Fig.~2 in][]{dwarfdm}. This effect is even more acute when photons from IC are accounted for -- in this case for large enough masses $m_{\rm DM}\gtrsim 100$ GeV, photons start to enter the {\em Fermi}-LAT energy range and to dominate the total photon count \citep[see Fig.~2 in][]{clusdm}. As a result of the combination of the flux limits improving with $m_{\rm DM}$ and of the photons counts decreasing with $m_{\rm DM}$, we expect a maximum, at some intermediate value of the particle mass, for the constraints on the lifetime. This indeed what we find, unless, as in the case of M31, we simply employ one fixed flux limit value.

In the $b\bar b$ panel, the yellow band to the left shows a parameter space range corresponding to theoretical predictions based on the gravitino dark matter model of \cite{2007JCAP...11..003B} and of \cite{2008PhRvL.100f1301I}, where we posit that the gravitino lifetime $\tau_{3/2}$ goes as
\begin{equation}
\tau_{3/2}\simeq3.8\times 10^{27} \ {\rm s}\left(\frac{U_{\rm grav}}{10^{-8}}\right)^{-2}\left(\frac{m_{3/2}}{10\ {\rm GeV}}\right)^{-3},
\end{equation}
with $m_{3/2}$ the gravitino mass, and the model-dependent parameter $U_{\rm grav}\sim10^{-8}$ (in the figure, the left boundary corresponds to $U_{\rm grav}=2\times10^{-8}$ and the right boundary to $U_{\rm grav}=0.5\times10^{-8}$). The right-most yellow band, in the $b\bar b$ panel, indicates a range for the mass-scale $M_{\rm NP}$ of a ``New Physics'' dimension-6 operator that would mediate dark matter decay in the context of supersymmetric grand unification, according to the rate outlined in Eq.~(\ref{eq:gut}) with $M_{\rm GUT}=M_{\rm NP}$. We take as an indicative range $10^{16}\lesssim M_{\rm NP}/{\rm GeV}\lesssim 4\times 10^{16}$.

In the right panel, we show decay lifetimes giving a good fit to the PAMELA positron excess and the combination of the PAMELA position fraction and the feature observed in the $e^+ + e^-$ spectrum measured by the {\em Fermi}-LAT \citep[for details on the determination of these contours, see][]{2010JCAP...03..014P}. We also shade in yellow the more ``optimistic'' limits calculated by \cite{2010JCAP...06..027Z}, where a specific, albeit rather arbitrary, Galactic gamma-ray diffuse model is assumed and subtracted off of the {\em Fermi} data.

It can be seen from Figure 1 that clusters of galaxies give significantly stronger constraints on the dark matter decay lifetime, both for a $b\bar b$ and a $\mu^+\mu^-$ final state and for a large range of particle masses, than have previously been derived from gamma-ray observations.  Specifically, for a $b\bar b$ final state the decay lifetime is limited to be above $\sim 5 \times 10^{26}$ secs for particle masses in the very broad mass range between 10 GeV and 1 TeV, while previous limits were about a factor of five worse worse. Limits on specific particle theories with meta-stable dark matter, such as those shown in the figure with shaded yellow contours, also improve accordingly.

For a $\mu^+\mu^-$ final state (right panel), the non-detection of clusters by {\em Fermi} excludes essentially the entire best fit region for dark matter decay models fitting the combination of the PAMELA positron fraction and the {\em Fermi}-LAT $e^+ + e^-$ spectrum and much of the region fitting the PAMELA data alone. Notice that this is in line with the predictions of \cite{pinzke}, where the effect of substructures was also extensively discussed.  Only models where the dark matter particle is significantly lighter than 2 TeV seem to pass the test of gamma ray emission from galaxy clusters. These results, therefore, cast doubt on the dark matter decay explanation of the rising positron fraction at high energies, particularly if this is taken jointly as an explanation of the {\em Fermi}-LAT electron-positron spectrum.  Similar to dark matter annihilation, the strongest constraints come from gamma-ray observations of nearby groups or poor clusters like Fornax, M49, and NGC4636, but all of the clusters in our sample lead to an improvement in the dark matter decay constraints. We note from Table 1 that the ranking in terms of the brightest signal from dark matter of the various clusters is quite different for the cases of decay and annihilation, although the brightest object is in both cases the Fornax cluster. Notice that some uncertainty is present in the determination of the coefficient $J_d$, as we discuss below.

In Figure \ref{fig:clussys}, we compare the constraints on the dark matter decay lifetime determined above (blue lines) using an X-ray derived dark matter density profile to alternate choices for the density profile or cluster mass as an indication of the potential systematic error in our constraints for the Fornax cluster (left panel), which has the highest $J_d$ of the cluster sample, and for the Coma cluster (right panel), which has the highest mass of the cluster sample.  We only explicitly show decay to $\mu^+\mu^-$, but the same relative offsets hold for other final states.  For the Fornax cluster, we plot the limits for two alternate estimates of the cluster mass: the mass from \cite{2001ApJ...548L.139D} based on the velocity dispersion of cluster member galaxies and the mass from the independent X-ray analysis of \cite{2002ApJ...565..883P} based on ROSAT HRI observations (mass within one degree given from their published mass profiles).  The dark matter density profile in the central Fornax galaxy, NGC1399, has also been investigated based on the radial velocities of globular clusters (GC), but GC dynamics (particularly the red GC) probe primarily the central galaxy itself and do not reliably give the potential of the cluster as a whole \citep{2010A&A...513A..52S}.  The mass of the Coma cluster, the most massive and most distant cluster in our sample, has also been estimated based on weak lensing of background galaxies.  In the right panel of Figure \ref{fig:clussys}, we show the change in the constraints on the dark matter decay lifetime for three different weak lensing estimates of the Coma cluster mass \citep{2007ApJ...671.1466K, 2009A&A...498L..33G, 2010ApJ...713..291O} as well as for the cluster mass estimated based on the velocity dispersion of the cluster galaxies \citep{2003MNRAS.343..401L}.  For both Fornax and Coma, the constraints for a variety of mass estimates change by less than a factor of two from the nominal constraints shown in Figure 1.  Even the most conservative Fornax constraints represent a significant improvement over previous limits and exclude large portions of the parameter space that could explain the cosmic ray anomalies.  For our nominal constraints we employ the mass-concentration relation of \cite{conc}, because this relation was derived based on X-ray cluster observations; however, we note that for a variety of mass-concentration relations from both simulations and observations \citep{2009ApJ...707..354Z,2008MNRAS.391.1940M,2010arXiv1002.3660K,2010arXiv1009.3266E} the $J_d$ of Fornax would change by less than 20\%.


A full accounting for systematic effects for all clusters in the sample is not statistically not feasible (the observations not being directly comparable), nor do we wish to debate the merits of one set of observations over another.  However, Figure~\ref{fig:clussys} leads us to conclude that our constraints are robust within roughly a factor of two, and that they are not systematically biased high or low.  As the Fermi non-detection of most of the clusters in the sample excludes  a significant fraction of the decay lifetime and particle mass combinations needed for a joint dark matter decay explanation to the cosmic-ray anomalies, our results lead to the conclusion that this possibility appears disfavored.

\subsection{DWARF SPHEROIDAL GALAXIES}\label{sec:dsph}

Figure \ref{fig:dwarfs} shows the limits on the dark matter decay lifetime for nearby dwarf spheroidal galaxies using the {\em Fermi} gamma-ray limits published in \cite{dwarfdm}.  For the dwarf constraints with a $\mu^+\mu^-$ final state (right panel), we conservatively only include gamma-ray emission from final state radiation and do not include IC emission, given the unknown effects of diffusion on $e^+$ and $e^-$ in dwarf scale systems \citep{profumojeltemadsph, dracoullio}.  Below we discuss how these constraints would improve with the inclusion of IC gamma-ray emission and specific assumptions for cosmic-ray diffusion.   Compared to clusters, the non-detection of dwarf spheroidal galaxies in gamma-rays gives weaker constraints at least for the conservative limits we derive assuming a mass of $M_{1/2}$ and no IC emission.  However, at low masses the constraints still improve upon previous limits from the literature, in particular those we derive from the Ursa Minor dwarf galaxy.  

If we relax our conservative assumptions on gamma-ray emission from high-energy muon pairs, the constraints shown in the right panel of Figure~\ref{fig:dwarfs} become considerably stronger, as shown in Figure \ref{fig:dwarfsys} for Ursa Minor (we employ this specific dwarf galaxy as it gives the strongest constraints among the dwarf sample).  In the left panel of Figure \ref{fig:dwarfsys}, we consider the limits on the dark matter decay lifetime if we include the expected IC gamma-ray emission for a $\mu^+\mu^-$ decay final state for different assumed diffusion coefficients.  Following \cite{dwarfdm}, the blue lines show the constraints for diffusion coefficients of $D_0 = 10^{28}$ cm$^2$ s$^{-1}$ (dot-dashed line) and $D_0 = 10^{29}$ cm$^2$ s$^{-1}$ (dashed line) which bound the values found for the Milky Way.  In both cases, the energy dependence of diffusion is assumed to have a power-law dependence given by $D(E)=D_0\ \left(\frac{E}{\rm 1\ GeV}\right)^{1/3}$.  The red line shows the constraints in the limit of no diffusion (i.e. in the limit where the electrons and positrons loose all of their energy via IC prior to diffusing away from the system).  The inclusion of IC emission significantly strengthens the constraints from Ursa Minor, but for reasonable diffusion coefficients these are still weaker than previous limits on the decay lifetime.

In addition for our nominal constraints in Figure \ref{fig:dwarfs}, we conservatively assumed dwarf masses equal to $M_{1/2}$ even though the dwarf half-light radii are smaller than 1 degree.  In the right panel of Figure \ref{fig:dwarfsys}, we compare the constraints for Ursa Minor for different assumptions on the mass profile.  For illustration we show the limits on the decay lifetime for a $\mu^+\mu^-$ final state with a diffusion coefficient of $D_0 = 10^{28}$ cm$^2$ s$^{-1}$, but of course the same relative offset also holds for other cases.  Employing the best-fit NFW profile parameters from fits to the stellar velocity data for Ursa Minor published in \cite{dwarfdm} to calculate $J_d$ for a radius of 1 degree yields the dot-dashed line in Figure \ref{fig:dwarfsys} (right panel), a factor of a few stronger than for $M_{1/2}$.  \cite{wolf} find that all of the Milky Way dwarf spheroidals are consistent with having the same total halo mass (within the scatter) of around $3 \times 10^{9}$ M$_{\odot}$ for a $\Lambda$CDM cosmology.  Looking for example at their Figure 3, for a range of two orders of magnitude in total halo mass between $3 \times 10^{8}$ M$_{\odot}$ and $3 \times 10^{10}$ M$_{\odot}$ and for an NFW profile consistent with a WMAP5 cosmology, would give a mass within 1 degree for Ursa Minor between a few times $10^7$ M$_{\odot}$ and a few times $10^8$ M$_{\odot}$. The decay constraints for this range of masses is reflected in the two dashed lines in Figure \ref{fig:dwarfsys}, where we show the results for the specific values of $M(r<1^{\circ}) = 3\times 10^7$ M$_{\odot}$ and $3\times 10^8M_\odot$.  Depending on the assumptions on mass and diffusion, the Ursa Minor constraints become comparable to or stronger than previous limits.

\subsection{M 31}\label{sec:m31}

Although faintly, the nearby spiral galaxy M31 is detected at gamma-ray frequencies, as recently reported by the {\em Fermi}-LAT Collaboration \citep{fermim31}. The detected gamma-ray emission is compatible with a point-source, although a slightly higher test-statistics was reported if a spatial template based on the infra-red surface brightness was employed \citep{fermim31}. Here, to set constraints on the emission from dark matter decay we simply consider the 2-$\sigma$ upper limit on the observed emission, conservatively summing the statistical and systematic uncertainties quoted by the {\em Fermi}-LAT Collaboration. A more in-depth study would resort to either attempting to subtract the astrophysical emission from M31 as expected from cosmic-ray interaction with the inter-stellar medium (in analogy to the diffuse gamma-ray emission from our own Milky Way), or to utilize, for instance, a region of interest that cuts out the gamma-ray emitting region (e.g.~an annulus around the M31 central region). We defer such a study for a forthcoming analysis.

M31 is an extremely interesting object to search for a signal from dark matter decay, as was for instance suggested, for the case of sterile neutrinos (producing a monochromatic line at X-ray frequencies from the two-body decay into an active neutrino and a photon) by \cite{sterilenu}. To estimate the object's brightness in gamma rays from dark matter decay, we employ the NFW dark matter density profile proposed by \cite{2006MNRAS.366..996G}: Table 1 shows that we obtain a $J_d$ value (1 degree radius) for M31 which exceeds by almost a factor 2 the largest normalization factor for galaxy clusters, and by more than an order of magnitude the largest one we find among dwarf galaxies.

Despite being a potentially extremely bright source of gamma rays from dark matter decay, the constraints we derive from M31 are not as strong as those we found for clusters -- the reason evidently being that M31 is actually detected as a gamma-ray source, and we conservatively do not attempt to subtract off any astrophysical background. The left panel of Figure~\ref{fig:m31} illustrates the constraints we obtain for a $b\bar b$ decay final state. The black line refers to the NFW profile with the gamma-ray flux obtained with the assumption of a point-like source and $J_d$ as listed in Table 1. The green dashed line assumes the isothermal profile of \cite{2001MNRAS.323...13K} and the same gamma-ray flux limit -- the closeness of the two lines illustrates the small uncertainty that stems from using two entirely different functional forms for the dark matter density profile. Finally, for the blue dotted line we employ the extended source {\em Fermi}-LAT flux and assume that the relevant mass coincides with the virial mass of M31 \citep[which we set to $M_{\rm 200}\simeq 6.8\times 10^{11}M_\odot$, from][]{2006MNRAS.366..996G}. The limits we obtain, while an improvement over what we get from dwarf galaxies and over previously available limits, are a factor of a few worse than those implied by the non-detection of galaxy clusters for the $b\bar b$ final state.

The right panel of Figure~\ref{fig:m31} shows our results for the $\mu^+\mu^-$ final state. We compare again with the regions favored by a dark matter decay interpretation of the cosmic-ray anomalies (green and red shaded regions) and with the current constraints (grey region). The black lines refer to the NFW plus point-like source assumption, while the red lines refer to the case where we employ the virial mass and the extended source gamma-ray flux. Given the relevance of diffusion for cosmic-ray electrons and positrons in the Milky Way, we do expect leakage of secondary $e^+e^-$ out of M31 as well, and we model this process again in analogy to our own Galaxy. We show with solid lines our results for the relatively small diffusion coefficient $D_0=10^{28}$ cm$^2$/s, and with dashed lines the more conservative case of $D_0=10^{29}$ cm$^2$/s (as for dwarf galaxies, we set the rigidity dependence power to 1/3). In either case, the limits we obtain do not improve over previous constraints nor do they approach the phenomenologically relevant regions for the cosmic-ray anomalies.

\section{DISCUSSION AND CONCLUSIONS}

In this paper, we analyzed how early {\em Fermi} limits on the gamma-ray flux from clusters of galaxies, from nearby dwarf galaxies and the {\em Fermi} detection of M31 impact the possible lifetime of a meta-stable particle dark matter. We showed that limits from dwarf galaxies and from M31 improve over previous constraints, for certain ranges of masses and final states. The best limits are however derived in the case of galaxy clusters, where we obtain an improvement of about a factor of five on the allowed decay lifetime for hadronic decay final states for a wide range of particle dark matter masses between 10 GeV and 1 TeV. Furthermore, models where dark matter decays into muon pairs and explains the so-called cosmic-ray anomalies (the positron excess reported by PAMELA and the feature observed by {\em Fermi} in the electron-positron spectrum) are found to over-produce gamma rays from clusters of galaxies compared to {\em Fermi} observations for large regions of the best-fitting parameter space. While we cannot completely exclude the possibility, our results imply that the dark matter decay interpretation of the cosmic-ray anomalies is in tension with gamma ray observations, especially if both the {\em Fermi} and PAMELA anomalies are jointly explained, and this scenario may be ruled out completely if Fermi does not detect gamma-ray emission from clusters in the next few years. {\em Fermi}-LAT continues to take data in all-sky survey mode, and in the future longer exposures and a better understanding of the astrophysical backgrounds will continue to improve constraints on dark matter models and perhaps even reveal a signal from this elusive component of the universe.

\acknowledgments
This work is partly supported by NASA grants NNX09AT96G and NNX09AT83G. SP also acknowledges support from the National Science Foundation, award PHY-0757911-001, and from an Outstanding Junior Investigator Award from the Department of Energy, DE-FG02-04ER41286. LD was supported by the UC LEADS Program.


\begin{deluxetable}{lccccccc}
\tablecaption{ Data for Clusters and M31 }
\tablewidth{0pt}
\tablecolumns{8}
\tablehead{
\colhead{Name} & \colhead{Distance} & \colhead{$M_{200}$} & \colhead{$c_{200}$} & \colhead{$R_{200}$} & \colhead{$r_{1 deg}/R_{vir}$} & \colhead{$J_d$} & \colhead{Ranking for}  \\
\colhead{} & \colhead{Mpc} & \colhead{$10^{14} M_{\odot}$} & \colhead{} & \colhead{Mpc} & \colhead{} & \colhead{$10^{18}$ GeV cm$^{-2}$} & \colhead{DM Annihilation}
}
\startdata
Fornax & 18.9 & 1.42 & 8.9 & 1.35 & 0.24 & 18.4 & 1\\
Coma & 94.9 & 19.38 & 5.7 & 3.22 & 0.50 & 16.6 & 5 \\
Centaurus & 42.2 & 3.78 & 7.5 & 1.87 & 0.39 & 13.7 & 4\\
M49 & 18.1 & 0.65 & 10.2 & 1.04 & 0.30 & 11.1 & 2 \\
AWM7 & 70.4 & 6.08 & 7.0 & 2.19 & 0.55 & 10.2 & 6 \\
NGC4636 & 15.2 & 0.35 & 11.4 & 0.85 & 0.31 & 8.88 & 3\\
\hline
M 31 & 0.77 & 0.0068 & 22.0 & 0.18 & 0.075 & 36.4 & -- \\
\enddata
\tablecomments{For clusters, $M_{200}$ and $R_{200}$ are taken from \cite{re02} and the concentrations are calculated from the concentration-mass relation of \cite{conc}; for M31 these quantities are taken from \cite{2006MNRAS.366..996G}.  Column 6 gives the ratio of the radius which subtends a projected angle of 1 degree (i.e. the radius within which $J_d$ is calculated) to $R_{200}$.  The $J_d$ factors, listed in column 7, for the clusters under consideration and for M31 are calculated as described in \S3.  For comparison, column 8 lists the relative ranking of the same objects in terms of their normalizations for a signal from dark matter annihilation.}
\end{deluxetable}

\clearpage

\begin{deluxetable}{lcccccc}
\tablecaption{ Data for Dwarf Spheroidal Galaxies }
\tablewidth{0pt}
\tablecolumns{7}
\tablehead{
\colhead{Name} &\colhead{Distance} & \colhead{$M_{1/2}$} & \colhead{$r_{1/2}$} & \colhead{$r_{1 deg}/r_{1/2}$} & \colhead{$J_d$} & \colhead{Ranking for}  \\
\colhead{} &  \colhead{kpc} & \colhead{$10^{7} M_{\odot}$} & \colhead{pc} & \colhead{} & \colhead{$10^{18}$ GeV cm$^{-2}$} & \colhead{DM Annihilation}
}
\startdata
Ursa Minor & 66 & 5.56 & 588 & 2.0 & 1.50 & 2 \\
Ursa Major II  & 30 & 0.79 & 184 & 2.8 & 1.03 & 3 \\
Bootes I   & 62 & 2.36 & 322 & 3.4 & 0.72 & 5 \\
Sextans     & 86 & 3.49 & 1019 & 1.5 & 0.55 & 7 \\
Fornax    & 138 & 7.39 & 944 & 2.6 & 0.45 & 8 \\
Draco & 76 & 2.11 & 291 & 4.6 & 0.43 & 1\\
Sculptor & 79 & 2.25 & 375 & 3.7 & 0.42 & 4 \\
Coma Berenices & 44 & 0.197 & 100 & 7.7 & 0.12 & 5 \\
\enddata
\tablecomments{The distances to the dwarf spheroidal galaxies are taken from \cite{dwarfdm}, while $M_{1/2}$ and $r_{1/2}$ are taken from \cite{wolf}.  Column 5 gives the ratio of the radius which subtends a projected angle of 1 degree (i.e. the radius within which $J_d$ is calculated) to $r_{1/2}$.  Column 6 lists the $J_d$ factors for dwarf galaxies under consideration calculated as described in \S3.  For comparison, column 7 lists the relative ranking of the same objects in terms of their normalizations for a signal from dark matter annihilation.}
\end{deluxetable}

\clearpage

\begin{figure}
\epsscale{0.5}
\mbox{\hspace*{-0.3cm}\plotone{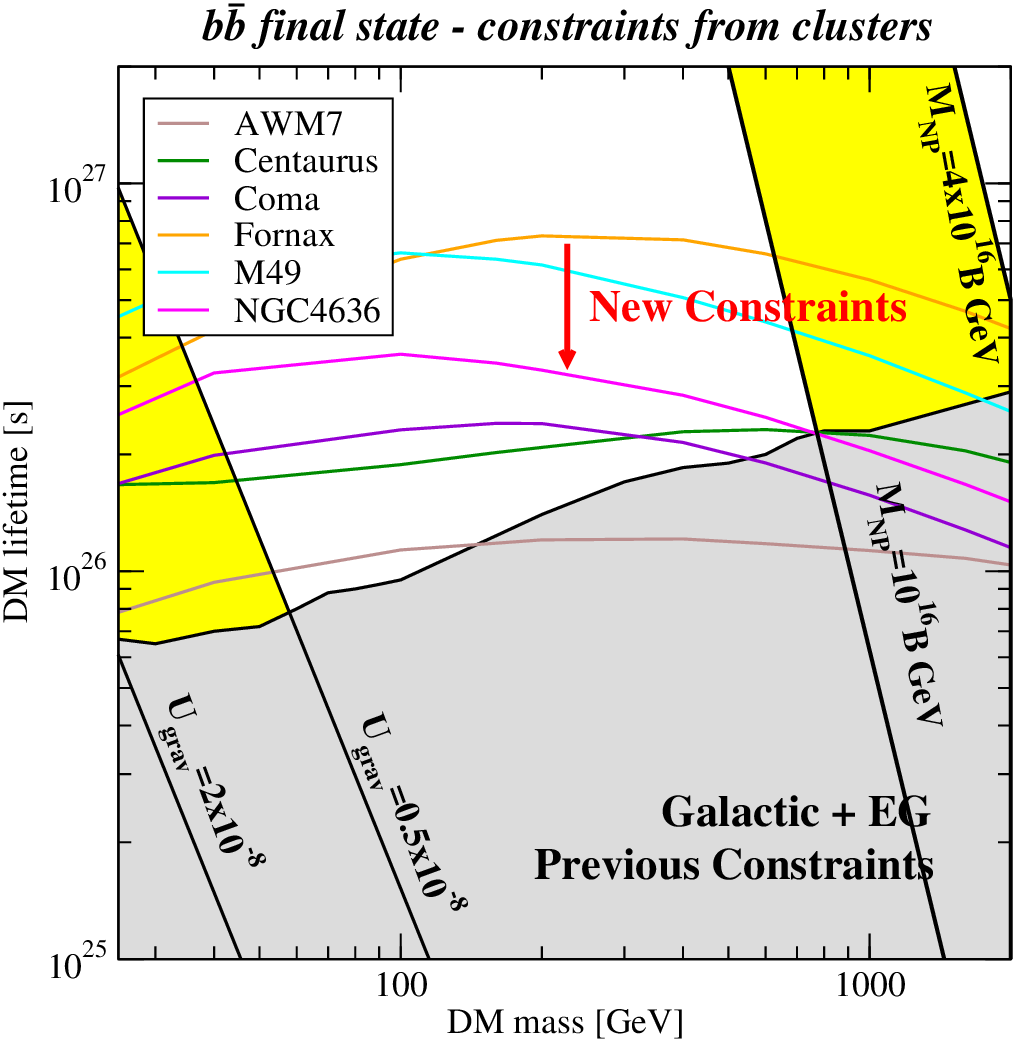}\quad
\plotone{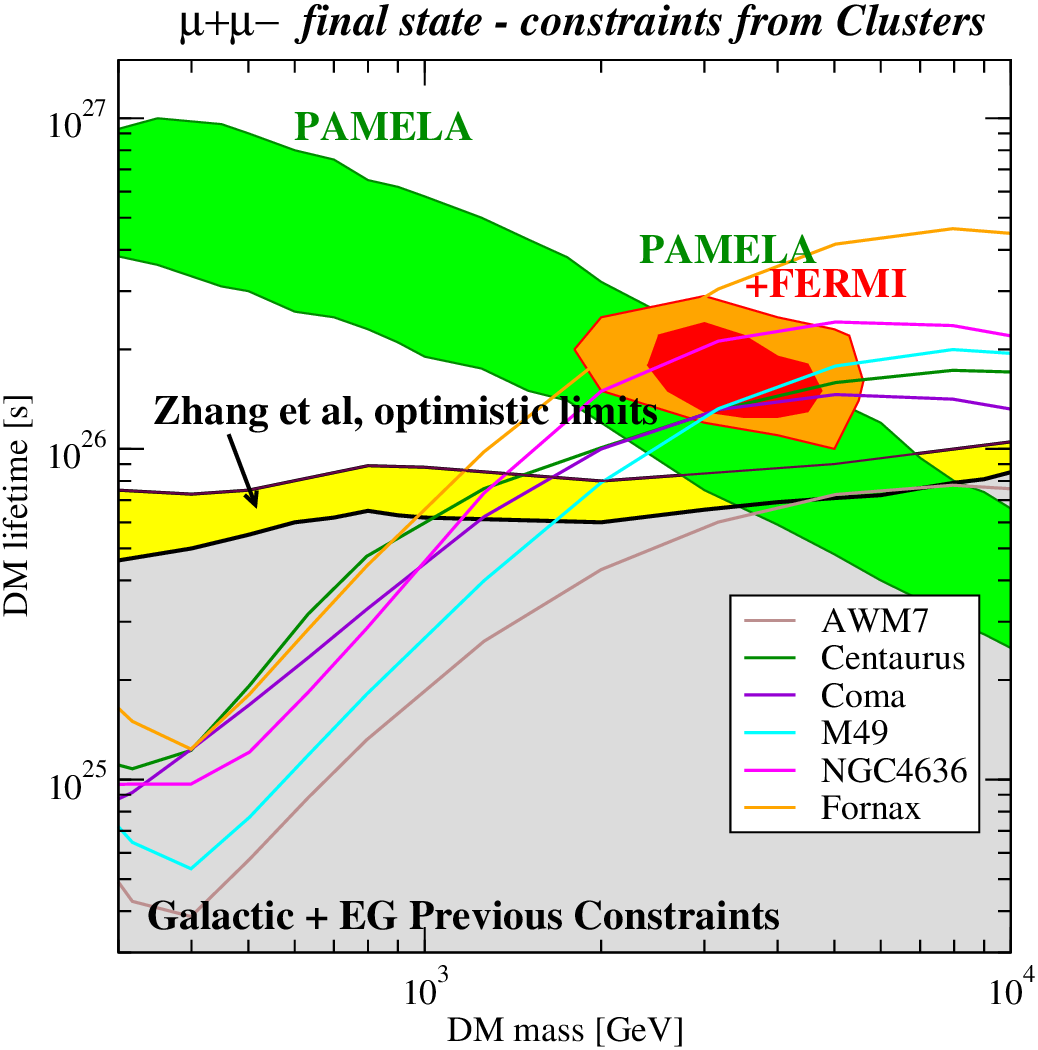}}
\caption{Constraints on the decay lifetime as a function of mass from clusters of galaxies for a $b\bar b$ final state (left panel) and a $\mu^+\mu^-$ final state (right panel) based on the 95\% confidence level upper limits on the gamma-ray flux from 11 months of {\em Fermi}-LAT observations. The grey shaded region shows an example of previous constraints from \cite{2010JCAP...06..027Z} which consider the expected diffuse dark matter decay signal from the Galactic halo and unresolved extragalactic dark matter; the yellow shaded region in the right panel shows the more optimistic limits from \cite{2010JCAP...06..027Z} obtained after subtraction of a model for the Galactic diffuse emission.  In the left panel, the yellow bands show the regions for theoretically motivated models for gravitino decay (leftmost yellow band) and for theories with a dimension-6 operator at the GUT scale mediating dark matter decay (rightmost yellow band), see \S4.1 for details.  In the right panel, we also show the regions of parameter space fitting the observed cosmic-ray anomolies \citep[green: PAMELA positron fraction only, red: the combination of the PAMELA position fraction and the feature observed in the $e^+ + e^-$ spectrum measured by the {\em Fermi}-LAT, see][]{2010JCAP...03..014P}. }
\end{figure}

\begin{figure}
\epsscale{0.5}
\mbox{\hspace*{-0.3cm}\plotone{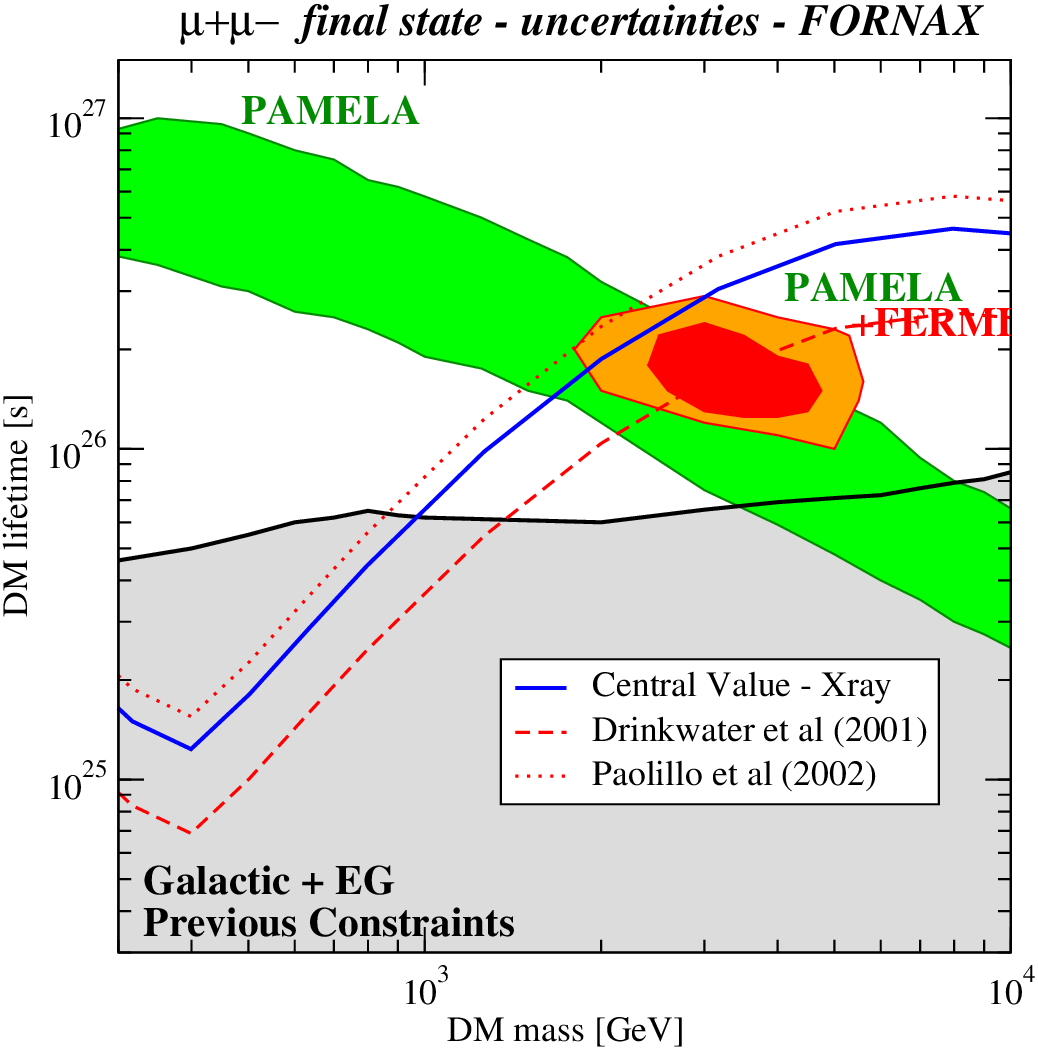}\quad
\plotone{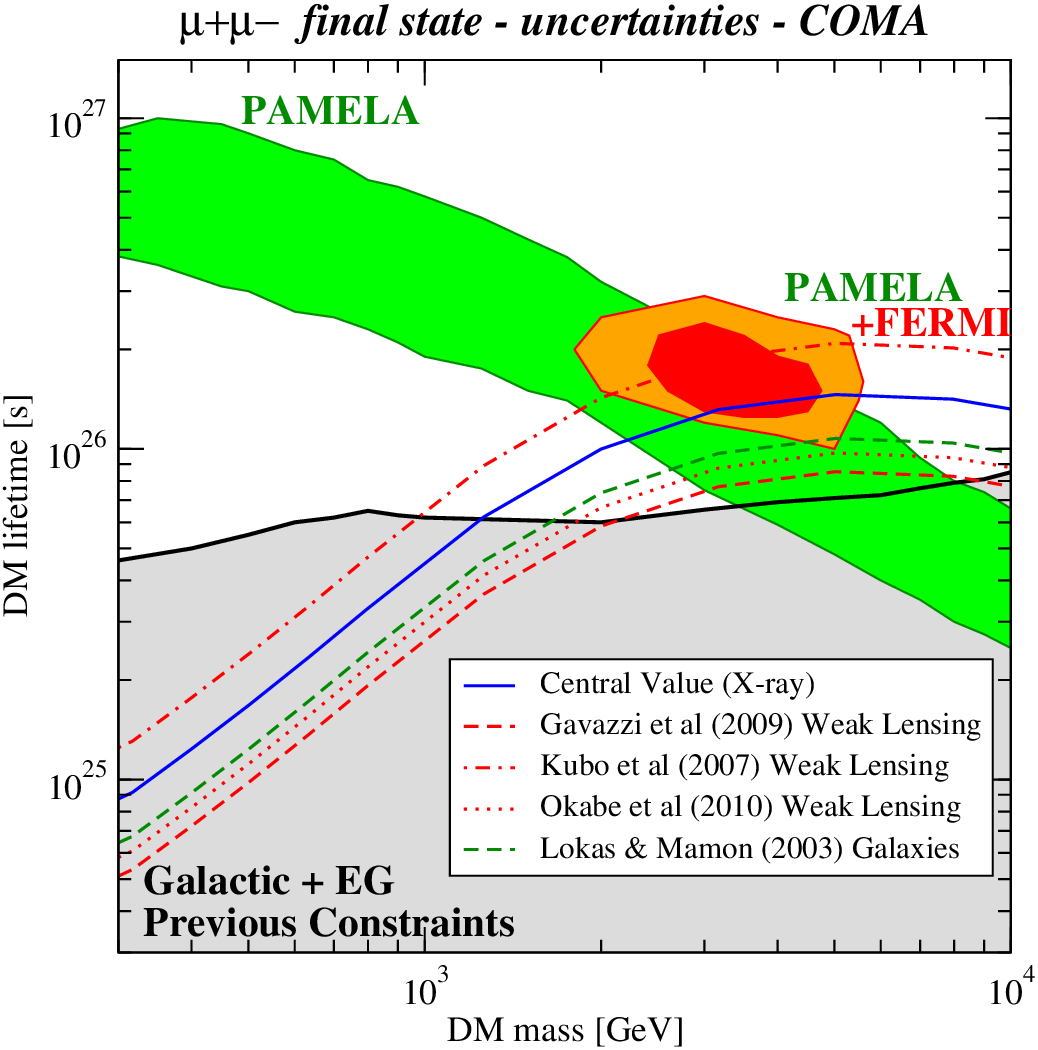}}
\caption{Constraints on dark matter decay lifetime as a function of particle mass for alternate determinations of the mass of the Fornax (left panel) and Coma (right panel) clusters.  The constraints are shown for decay to $\mu^+\mu^-$, but the same relative offsets hold for other final states.  The nominal constraints based on the X-ray determined masses within 1 degree are reproduced from Figure 1 by the solid blue lines (see \S3.1).  \textit{Left:} The red lines show the constraints if we employ, instead, the dark matter mass of Fornax (radius of 1 degree) determined from the velocity dispersion of cluster member galaxies \citep[red dashed line,][]{2001ApJ...548L.139D} or the independent, non-parametric X-ray analysis of \cite{2002ApJ...565..883P} (red dotted line).  
\textit{Right:} The red lines show the constraints on the dark matter decay lifetime for three different weak lensing estimates of the Coma cluster mass \citep{2007ApJ...671.1466K, 2009A&A...498L..33G, 2010ApJ...713..291O} while the green dashed line shows constraints if the mass is estimated based on the velocity dispersion of the cluster galaxies \citep{2003MNRAS.343..401L}. }
\label{fig:clussys}
\end{figure}

\begin{figure}
\epsscale{0.5}
\mbox{\hspace*{-0.3cm}\plotone{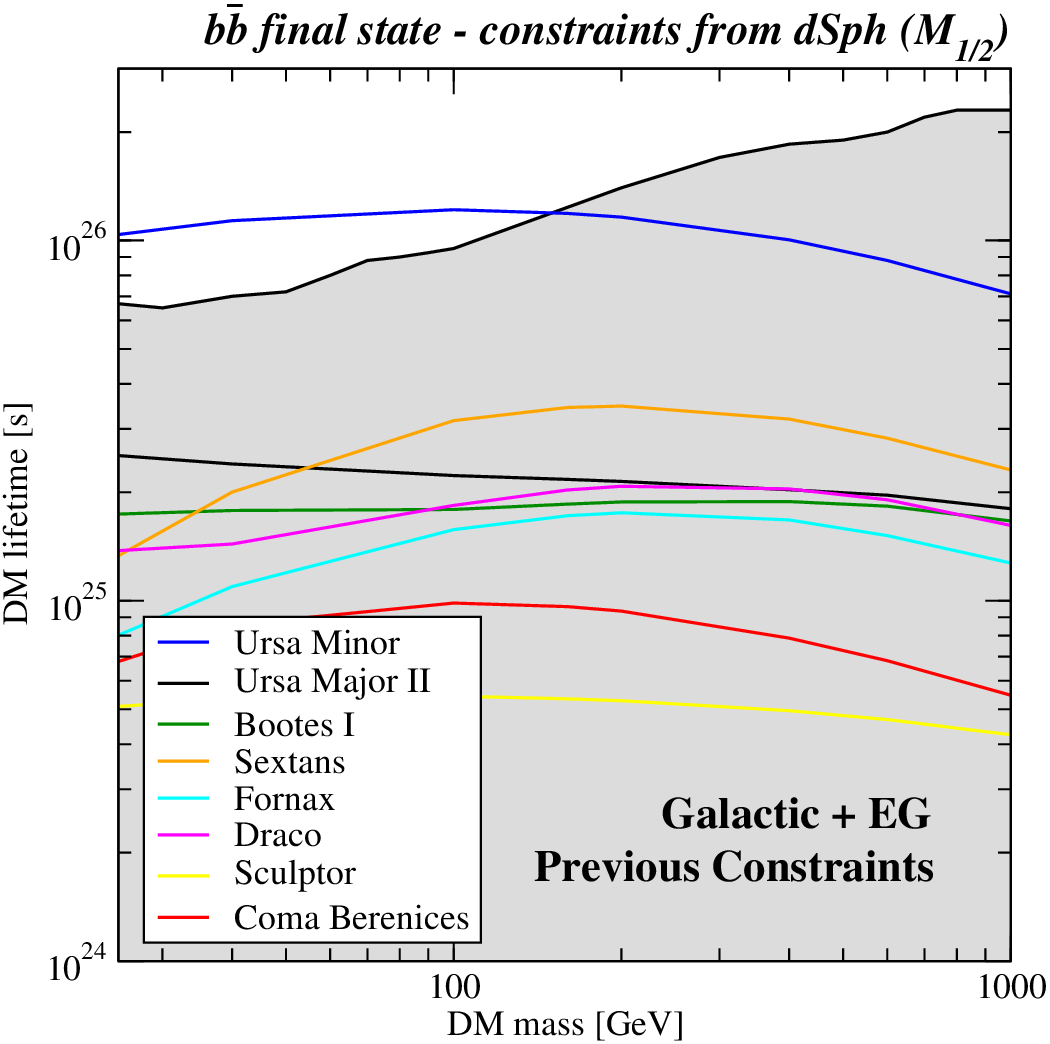}\quad
\plotone{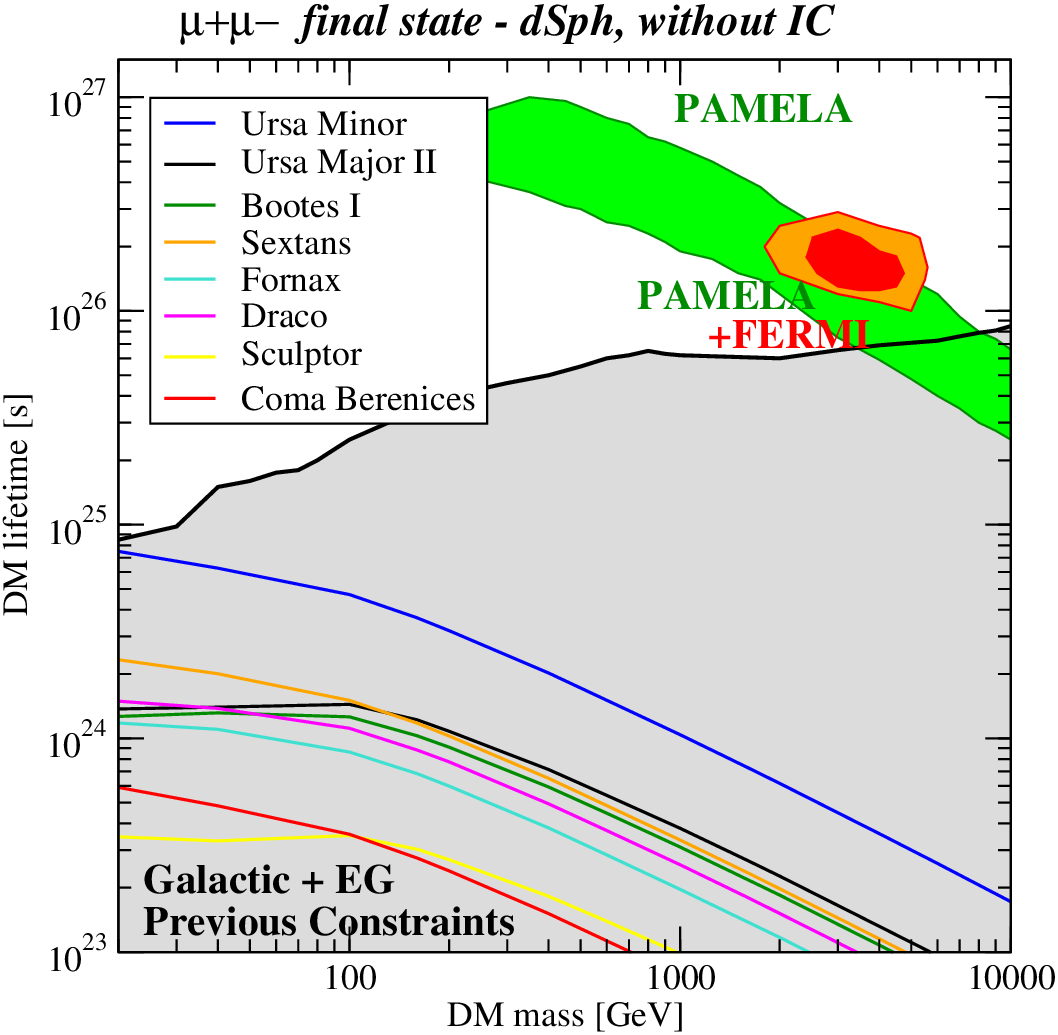}}
\caption{Constraints on the decay lifetime as a function of mass from dwarf spheroidal galaxies for a $b\bar b$ final state (left panel) and a $\mu^+\mu^-$ final state (right panel) based on the 95\% confidence level upper limits on the gamma-ray flux from 11 months of {\em Fermi}-LAT observations.  Here we conservatively assume no IC gamma-ray emission.}
\label{fig:dwarfs}
\end{figure}

\begin{figure}
\epsscale{0.5}
\mbox{\hspace*{-0.3cm}\plotone{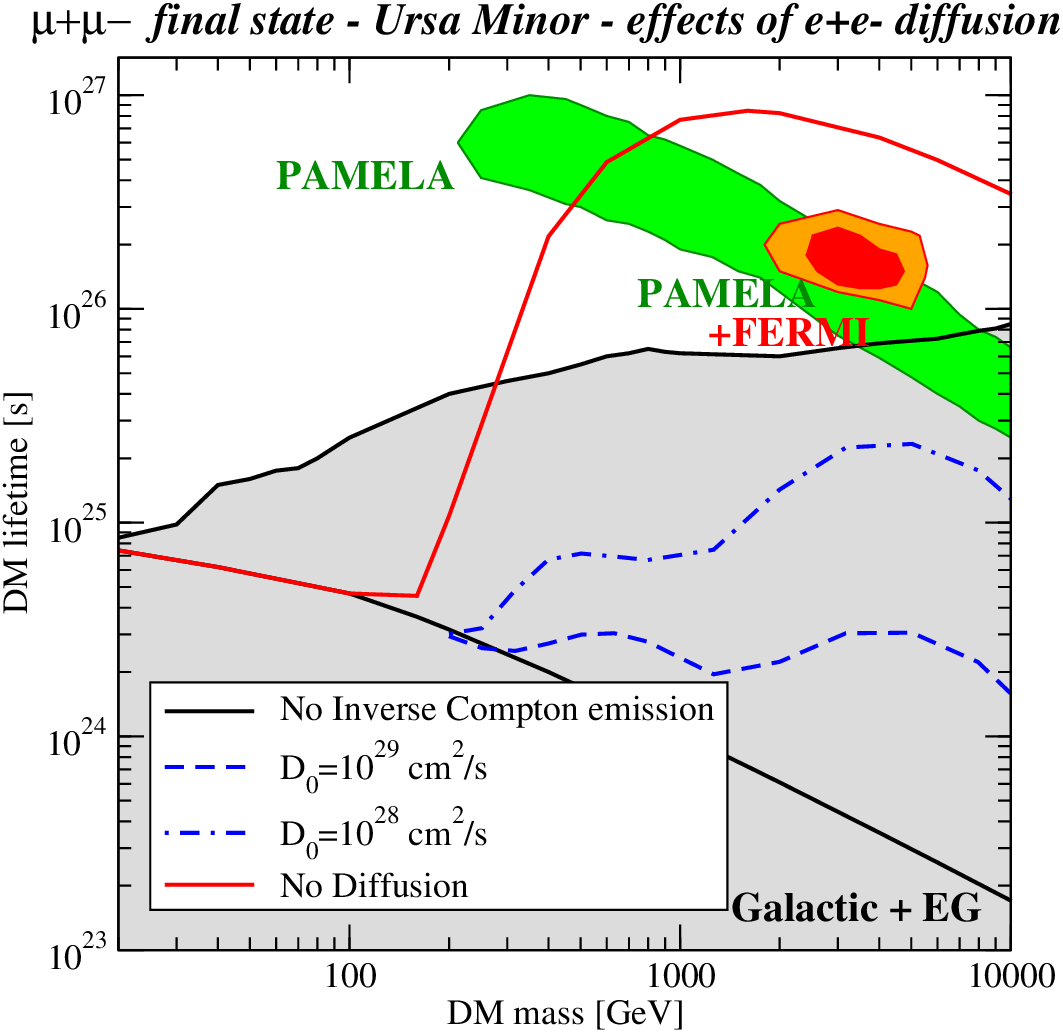}
\plotone{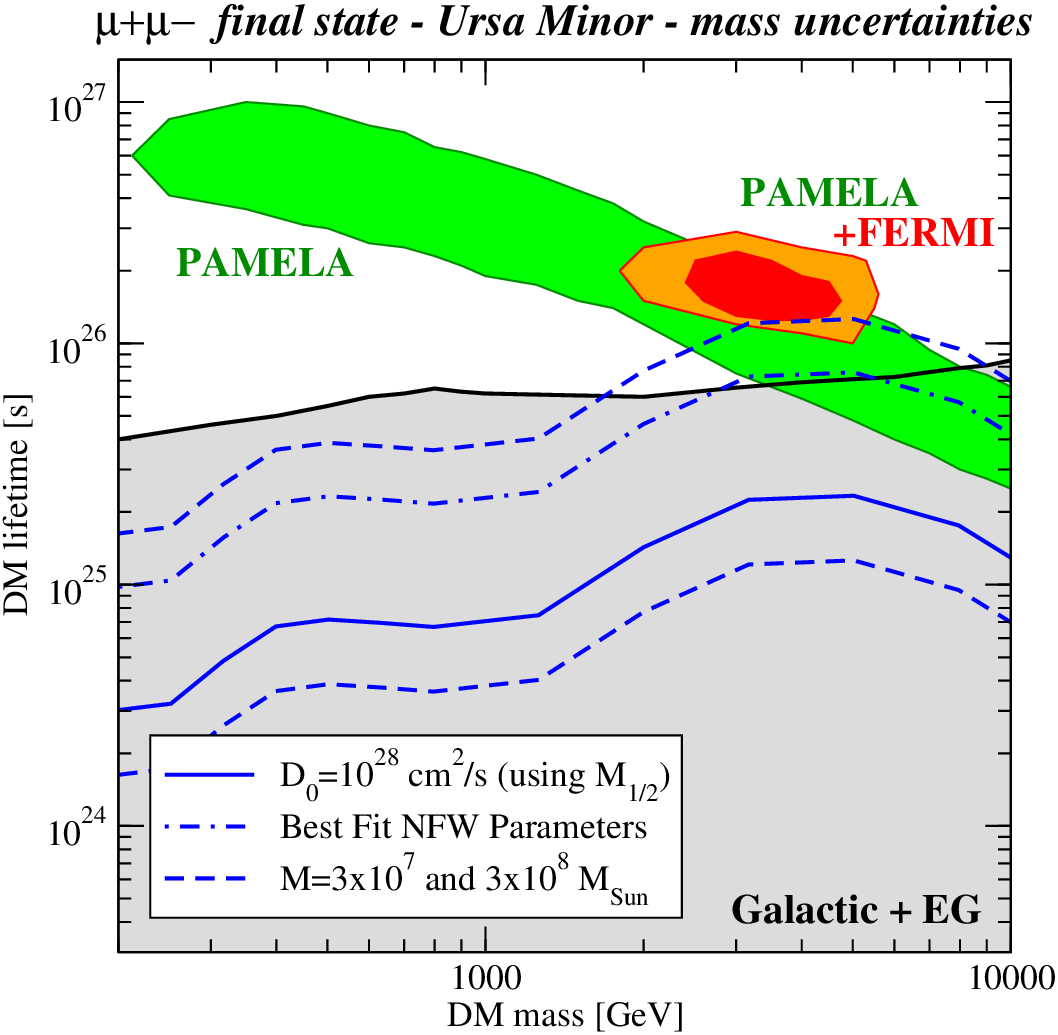}}
\caption{Constraints on the dark matter decay lifetime as a function of particle mass for the Ursa Minor dwarf spheroidal galaxy for an assumed $\mu^+\mu^-$ final state. \textit{Left:} Constraints including IC emission for different choices of the diffusion coefficient $D_0$.  The black line conservatively shows the constraints for no IC emission, the blue lines include IC emission for two diffusion coefficients ($D_0 = 10^{28}$ cm$^2$ s$^{-1}$: dot-dashed line and $D_0 = 10^{29}$ cm$^2$ s$^{-1}$: dashed line) bounding those derived for the Milky Way and similar to what was assumed for dark matter annihilation in \cite{dwarfdm}, and the red line shows the constraints in the limit of no diffusion.  \textit{Right:}  Constraints for a diffusion coefficient of $D_0 = 10^{28}$ cm$^2$ s$^{-1}$ for different assumed masses for Ursa Minor. The solid line reproduces the constraints using the conservative assumption of a mass corresponding to the half-light radius, $M_{1/2}$.  For comparison, we also show the constraints using the NFW profile parameters from \cite{dwarfdm} based on fits to the Ursa Minor stellar velocity data and calculating $J_d$ within a radius of 1 degree (dot-dashed line).  The dashed lines show constraints on the decay lifetime for an order of magnitude variation in the assumed mass; these masses bracket the expected mass within a radius of 1 degree for dark mater density profiles fitting the stellar data for Milky Way dwarf spheroidal galaxies which are all consistent with having the same total halo mass \citep[][see for example their Figure 3]{wolf}.
\label{fig:dwarfsys}}
\end{figure}

\begin{figure}
\epsscale{0.5}
\mbox{\hspace*{-0.3cm}\plotone{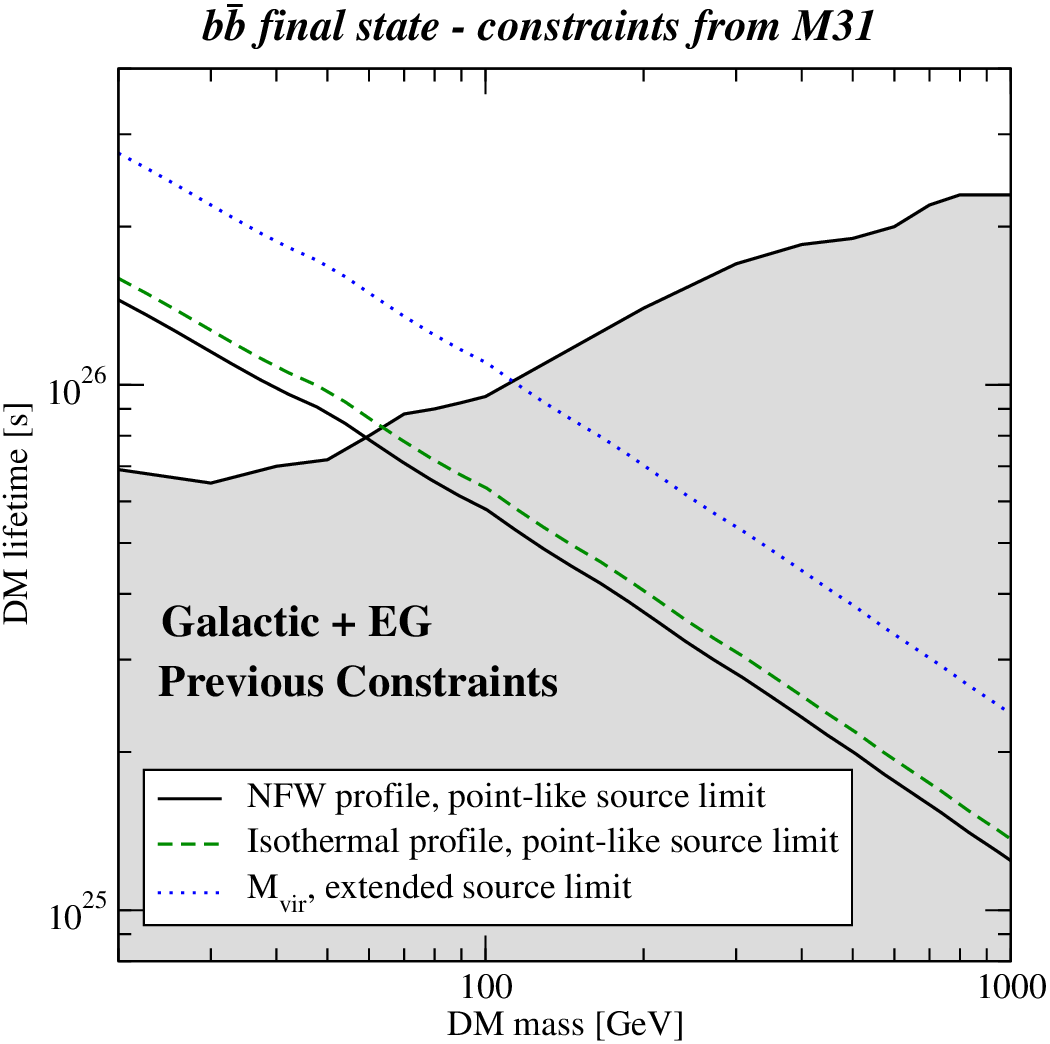}
\plotone{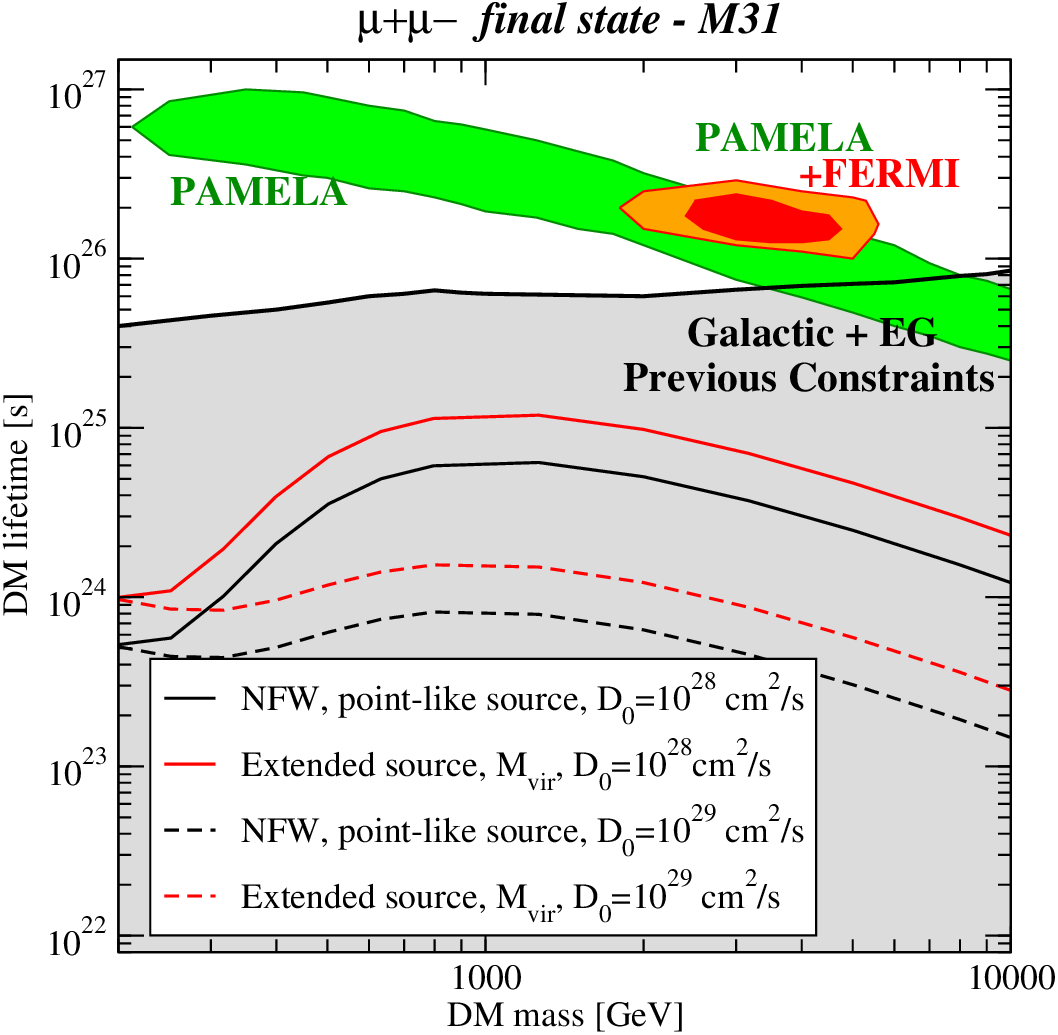}}
\caption{Constraints on the dark matter decay lifetime as a function of particle mass for the M31 galaxy, for a $b\bar b$ final state (left) and for a $\mu^+\mu^-$ final state (right). In the left panel, we show with a black line the constraints obtained employing a NFW profile with parameters from \cite{2006MNRAS.366..996G} (assuming a 1 degree projected radius) and the 2-$\sigma$ upper limit to the gamma-ray emission with an assumed point-like source spatial distribution; the green dashed line uses the same gamma-ray limit, but assumes the iso-thermal profile of \cite{2001MNRAS.323...13K}. Finally, for the blue dotted line we employ the M31 virial mass and the {\em Fermi} 2-$\sigma$ upper limit to the gamma-ray emission with an assumed extended source distribution. In the right panel, we show the constraints on a $\mu^+\mu^-$ final state, for diffusion coefficients set to $D_0=10^{28}$ cm$^2$/s (solid lines) and $D_0=10^{29}$ cm$^2$/s (dashed lines); the black lines refer to the NFW plus point-like source assumptions, while the red lines to the case where we employ the virial mass and the extended source gamma-ray flux.}
\label{fig:m31}
\end{figure}

\end{document}